\documentclass[twocolumn,twoside,5p,preprint]{elsarticle}
\biboptions{sort&compress}

\usepackage[utf8]{inputenc}
\usepackage[english]{babel}
\usepackage{amssymb,amsthm,amsmath,amstext,amsbsy,amsopn}
\usepackage{bbm}
\usepackage{graphicx}
\usepackage{isotope}
\usepackage{float}
\usepackage[dvipsnames]{xcolor}

\usepackage{hyperref}
\hypersetup{
    colorlinks = true,
    citecolor  = blue,
    linkcolor  = blue,
    urlcolor  = black
}

\definecolor{plot1}{rgb}{0.86, 0.08, 0.24}
\definecolor{plot2}{rgb}{0.25, 0.41, 0.88}
\definecolor{plot3}{rgb}{1.0, 0.55, 0}
\definecolor{plot4}{RGB}{61,153,86}

\newcommand{\la}{\langle}
\newcommand{\ra}{\rangle}
\newcommand{\Xmax}[1]{\ensuremath{#1_\text{max}}}

\newcommand{\ECstar}[1]{\text{EC}$_{#1}$}

\newcommand{\Nec}{N_\text{EC}}
\newcommand{\Mec}{\mathcal{M}^\text{EC}}
\newcommand{\Mci}{\mathcal{M}^\text{CI}}
\newcommand{\Mecpt}{\mathcal{M}^\text{PT}}

\begin{document}

\allowdisplaybreaks

\title{Excited states from eigenvector continuation: the anharmonic oscillator}

\author[tud]{M.~Companys Franzke}
\ead{margarida.companys@stud.tu-darmstadt.de}
\author[tud,emmi,mpik]{A.~Tichai}
\ead{alexander.tichai@physik.tu-darmstadt.de} 
\author[tud,emmi]{K.~Hebeler}
\ead{kai.hebeler@physik.tu-darmstadt.de}
\author[tud,emmi,mpik]{A.~Schwenk}
\ead{schwenk@physik.tu-darmstadt.de}

\address[tud]{Technische Universit\"at Darmstadt, Department of Physics, 64289 Darmstadt, Germany}
\address[emmi]{ExtreMe Matter Institute EMMI, GSI Helmholtzzentrum f\"ur Schwerionenforschung GmbH, 64291 Darmstadt, Germany}
\address[mpik]{Max-Planck-Institut f\"ur Kernphysik, 69117 Heidelberg, Germany}

\begin{abstract}
Eigenvector continuation (EC) has recently attracted a lot attention in nuclear structure and reactions as a variational resummation tool for many-body expansions.
While previous applications focused on ground-state energies, excited states can be accessed on equal footing. 
This work is dedicated to a detailed understanding of the emergence of excited states from the eigenvector continuation approach.
For numerical applications the one-dimensional quartic anharmonic oscillator is investigated, which represents a strongly non-perturbative quantum system where the use of standard perturbation techniques break down.
We discuss how different choices for the construction of the EC manifold affect the quality of the EC resummation and investigate in detail the results from EC for excited states compared to results from a full diagonalization as a function of the basis-space size.
\end{abstract}


\maketitle

\section{Introduction}

The solution of the nuclear many-body problem has seen significant progress over the last years, enabling a first-principles microscopic description based on nuclear Hamiltonians from chiral effective field theory combined with \emph{ab intio} many-body approaches~\cite{Herg20review,Hebe203NF}.
While the exact solution can be obtained from variational techniques such as configuration interaction (CI) or quantum Monte-Carlo (QMC) approaches, the exponential scaling of the underlying methods prevents its use in medium-mass systems.
As an alternative, a diverse toolbox of many-body techniques have been designed that expand the exact solution around a suitably chosen $A$-body reference state at low polynomial cost at the price of sacrificing the variational character, e.g., many-body perturbation theory (MBPT)~\cite{Holt14Ca,Tich16HFMBPT,Tichai2020review}, the in-medium similarity renormalization group (IMSRG)~\cite{Herg16PR,Stroberg2019,Heinz2020}, coupled-cluster (CC) theory~\cite{Hage14RPP,Bind14CCheavy,Novario2020a}, or self-consistent Green's function (SCGF) theory~\cite{Soma14GGF2N3N,Soma20SCGF,Arthuis2020a}.
These developments allowed \emph{ab initio} studies of atomic nuclei containing up to one hundred particles~\cite{Morr17Tin,Arthuis2020a,Miyagi2021} as well as global calculations of the nuclear chart up to the iron region~\cite{Stroberg2021}.

Even with powerful non-perturbative methods at hand, MBPT methods have undergone a major revival (see Ref.~\cite{Tichai2020review}) in nuclear structure theory due to the development of RG evolution techniques that make the nuclear many-body problem computationally more tractable by softening nuclear forces to lower resolution~\cite{Bogn07SRG,Bogn10PPNP}.
Still, the convergence of perturbative expansions is not guaranteed and a reliable extraction of nuclear observables is additionally complicated in quasi-degenerate many-body systems, such as open-shell nuclei.

In the past, various resummation techniques were developed that allow to extract observables from a possibly divergent expansion.
While such schemes have been applied successfully, their power usually relies on an \emph{a priori} knowledge of the asymptotic behavior of the expansion, information that is rarely available in realistic applications.
Consequently, practitioners desire a robust yet accurate tool that enables for extracting physical observables in a numerically reliable way.
Recently, eigenvector continuation (EC) has been introduced for systems where the Hamiltonian $H(c)$ admits a smooth dependence on some external parameter $c \in \mathbb{R}$.
The EC method is based on analytical continuation of the expansion outside its initial domain of convergence by performing several re-expansions of the Taylor series, thus, effectively shifting the reference point of the many-body expansion.
This allows to explore the system at coupling values that are outside of the initial domain of convergence~\cite{Frame2018,Demol2020EC,Demol20BMBPT,Avik2021} (see also Ref.~\cite{Mihalka2017resum} for a similar resummation approach developed in quantum chemistry).
The EC method has been applied in various many-body methods, but mainly restricted to the evaluation of ground-state properties.
Moreover, EC has been used as a powerful and accurate emulator for quantifying theoretical uncertainties due to the nuclear Hamiltonian~\cite{Koenig2020,Ekstroem2019,Furnstahl2020,Melendez2021EC,Yoshida2021ec}.

In this Letter, we investigate low-lying excited states in the EC framework. 
We consider the anharmonic oscillator as a strongly coupled benchmark system that is outside the range of applicability of standard perturbation theory calculations~\cite{Bender1969}, and in
all cases the low-lying spectrum can be benchmarked against the exact solution obtained from full diagonalization.
We show that EC targeted at the ground state is able to access the excited states as well, and that this can be further improved by including in the EC information from perturbative corrections for the excited states.

\section{Eigenvector continuation}
\label{sec:ec}

The EC method is used to target physical systems that smoothly depend on an external parameter $H(c)$.
In many applications there exists a regime $0 \leq c \leq c_\text{crit} < 1$, where the many-body problem is easier to solve than for the target value $c=1$.
While physical observables, e.g., energy eigenvalues, can change significantly when $c$ is varied, the eigenstates themselves often are less sensitive and remain in a low-dimensional manifold of the many-body Hilbert space when changing $c$ to its physical value $c=1$.

Consequently, the EC framework is performed in two successive steps. 
First a reference manifold containing $\Nec$ auxiliary states is constructed,
\begin{align}
    \Mec \equiv \{ | \Psi(c_i) \ra \, : \, i=1,...,\Nec \} \, ,
    \label{eq:ECman}
\end{align}
containing (approximate) eigenstates of the set of Hamiltonians $\{ H(c_i) \}$.
Subsequently, the Hamiltonian at physical coupling $H = H(1)$ is diagonalized within the manifold $\Mec$.
Since the auxiliary states are non-orthogonal, this gives rise to a generalized eigenvalue problem with Hamiltonian kernel $H_{ij} = \la \Psi (c_i) | H | \Psi(c_j) \ra$ and norm kernel $N_{ij} = \la \Psi (c_i) | \Psi(c_j) \ra$.

The construction of the EC manifold $\Mec$ in Eq.~\eqref{eq:ECman} does not assume specific properties of the defining many-body wave functions.
Following the previous work in Refs.~\cite{Demol2020EC,Demol20BMBPT}, $\Mec$ is here constructed from PT wave functions for both ground and excited states on top of a non-degenerate reference state.

Perturbation theory starts from a partitioning of the initial Hamiltonian $H = H_0 + H_1$ into an unperturbed part $H_0$ and a perturbation $H_1$.
The 0-th order reference state is an eigenstate of the unperturbed Hamiltonian
\begin{align}
    H_0 |\Phi_n^{(0)} \ra = E_n^{(0)} |\Phi_n^{(0)} \ra \, ,
\end{align}
where the subscript $n$ labels the ground and excited states.

Introducing a parameter-dependent Hamiltonian $H(c) \equiv H_0 + c H_1$
the exact many-body wave function is written as an infinite Taylor expansion in terms of an auxiliary parameter $c$,
\begin{align}
    | \Psi_n (c) \rangle &= \sum_{p=0}^\infty c^p |\Phi_n^{(p)}\rangle  \, ,
\end{align}
where $|\Phi_n^{(p)}\ra$ denotes the $p$-th order state correction.

Using a perturbative ansatz for the many-body wave functions the EC manifold can be re-expressed via the transformation
\begin{align}
\begin{pmatrix}
 | \Psi_n (c_1) \rangle \\
 | \Psi_n (c_2) \rangle \\
 \vdots \\
 | \Psi_n (c_{\Nec}) \rangle
 \end{pmatrix}
 =
 \begin{pmatrix}
1 &c_1  &\cdots &c_1^P \\
 1 &c_2 &\cdots &c_2^P \\
 \vdots &\vdots &\ddots &\vdots \\
 1 &c_{\Nec}   &\cdots &c_{\Nec}^P
 \end{pmatrix}
 \begin{pmatrix}
 | \Phi^{(0)}_n \rangle \\
 | \Phi^{(1)}_n \rangle \\
 \vdots \\
 | \Phi^{(P)}_n \rangle
\end{pmatrix},
\end{align}
thus, yielding an equivalent representation of the EC manifold
\begin{align}
    \Mecpt_n \equiv \{ |\Phi^{(p)}_n\ra \, : \, p=1,...,P \} \, ,
\end{align}
where $P$ is the maximum perturbation order considered in the wave function~\cite{Demol2020EC,Demol20BMBPT}.
Finally, the explicit set of parameters $\{ c_i \}$ does not enter in this setup and the dimension of $\Mec$ is set by the maximum perturbative order $P$.

Consequently, the EC approach amounts to the numerical solution of a generalized eigenvalue problem~\cite{Frame2018,Demol2020EC,Demol20BMBPT},
\begin{align}
    \mathbf{H} X = E \mathbf{N} X \, ,
    \label{eq:geneval}
\end{align}
where $\mathbf{H}$ and $\mathbf{N}$ are the Hamiltonian and norm matrices expanded in the basis of PT state corrections, 
\begin{subequations}
\begin{align}
    \mathbf{H}_{pq} &\equiv \langle \Phi_n^{(p)} | H | \Phi_n^{(q)} \rangle \, , \\
    \mathbf{N}_{pq} &\equiv \langle \Phi_n^{(p)} | \Phi_n^{(q)} \rangle \, .
\end{align}
\label{eq:ECpt}%
\end{subequations}
The quantity $E$ denotes the $P+1$ generalized eigenvalues and corresponds to the EC energies.
By employing intermediate normalization, i.e., $\la \Phi^{(0)}_n| \Psi \ra = 1$ with all basis states, the norm matrix fulfils $\mathbf{N}_{0p} = \delta_{0p}$ and
the Hamiltonian matrix entries within the first row (and column) correspond to the MBPT energy corrections, i.e., $\mathbf{H}_{0p} = \mathbf{H}_{p0} = E^{(p+1)}$.
However, when $p,q\neq0$ the matrix element $\mathbf{H}_{pq} \sim c^{p+q}$ contains many-body correlations up to order $p+q$, thus, going beyond a simple PT evaluation.
Due to the final diagonalization the EC approach is intrinsically non-perturbative and resums correlations not present in a simple PT approach. 
Moreover, the diagonalization ensures the EC framework to be manifestly variational, such that going to higher orders guarantees an improvement in accuracy as opposed to most medium-mass many-body frameworks applicable.

In practice, high-order energy and state corrections can be accessed using a recursive formulation of PT~\cite{Roth10PadePT,Tich16HFMBPT}.
Computationally, all quantities are obtained from matrix-vector multiplications of the interacting Hamiltonian represented in the unperturbed basis, i.e., in terms of eigenstates of $H_0$.

For targeting excited states, we investigate two different strategies for EC. First, perturbative state corrections are evaluated for the many-body ground state and excited states are accessed as eigenvalues from the solution of the generalized eigenvalue problem by forming Hamilton and norm matrices in Eq.~\eqref{eq:ECpt} with $n=0$.
Second, the acronym \ECstar{n} is used to indicate that the PT state corrections are evaluated for the $n$-th excited state in the unperturbed spectrum $|\Phi^{(p)}_n\ra$ and the EC matrices are constructed from these states. For \ECstar{0}, this coincides with the first strategy.

\section{Anharmonic oscillator}
\label{sec:aHO}

As an example we consider the one-dimensional harmonic oscillator with a quartic anharmonic term
\begin{align}
    H_\text{aHO}(c) \equiv \frac{1}{2m} p^2 + \frac{1}{2}m \omega x^2 +  c x^4 \, ,
\end{align}
where $m$ is the mass of the particle, $\omega$ the oscillator frequency, and $p$ and $x$ the momentum and position operators, respectively.
The strength of the perturbation is controlled by $c$. In the following, we work in natural units with $m$ and $\omega$ set to unity.
For the perturbative treatment the anharmonic oscillator (aHO) Hamiltonian is partitioned according to
\begin{subequations}
\begin{align}
    H_0 &= \frac{1}{2} p^2 + \frac{1}{2} x^2 \, , \label{eq:H0} \\
    H_1 &= x^4 \, , \label{eq:H1}
\end{align}
\end{subequations}
where the unperturbed system corresponds to the harmonic oscillator (HO) that can be solved exactly, with an equidistant spectrum $\epsilon_n  = n +1/2$ with quantum number $n=0, 1, 2,\ldots$.
In the following $\Xmax{n}$ defines the maximum excitations $n \leq \Xmax{n}$ in the model space.

\begin{figure}[t]
    \centering
    \includegraphics[width=1.\columnwidth]{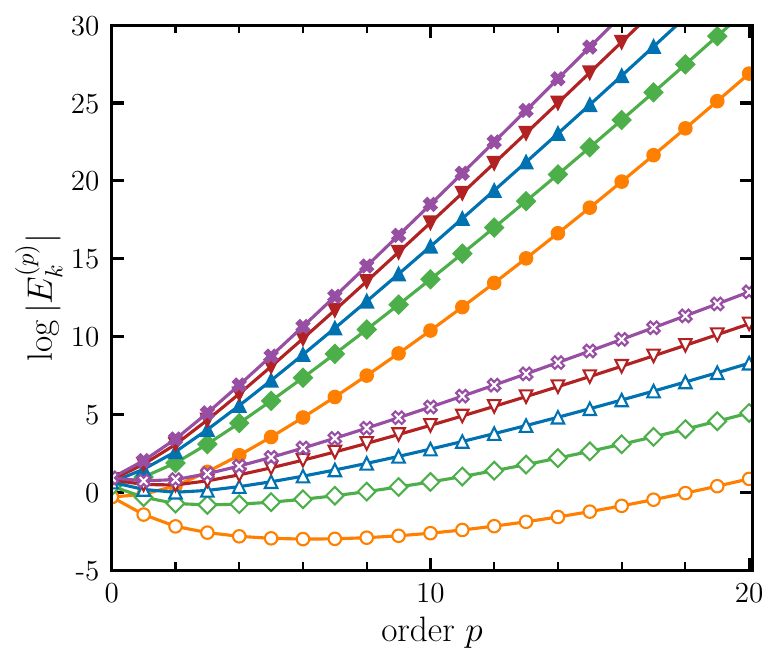}
    \caption{Absolute value of the energy corrections as a function of perturbation order for the five lowest states in the positive-parity spectrum of the anharmonic oscillator. Closed symbols correspond to $c=1$ whereas open symbols correspond to $c=0.1$. The model-space dimension is given by $\Xmax{n}=50$.}
    \label{fig:PTdiv}
\end{figure}

The seminal work by Bender and Wu revealed that the application of perturbation theory for the aHO Hamiltonian leads only to an asymptotic expansion for arbitrary coupling strength due to the presence of a branch-cut singularity at the origin~\cite{Bender1969}.
The divergence of the energy corrections for different reference states is shown in Fig.~\ref{fig:PTdiv}. It is clear that the energy corrections are exponentially growing, thus, making an extraction of observables from the bare perturbation series impossible.

Consequently, the aHO Hamiltonian provides a non-trivial testcase to study the low-lying spectrum within the EC framework.
We note that the spectrum of the unperturbed Hamiltonian is non-degenerate such that excited states can be targeted on equal footing without resorting to (quasi-)degenerate PT extensions that further require a diagonalization within a degenerate subspace.
The reference states for the recursive PT evaluation are chosen as $|\Phi^{(0)}_n \ra  = | n \ra$, i.e., as excited eigenstates of the unperturbed HO [Eq.~\eqref{eq:H0}].

Since the aHO Hamiltonian is parity-conserving, eigenstates with different parities $\Pi$ do not mix, and we can therefore consider the Hamiltonian and EC for positive and negative parity separately. For completeness, matrix elements of the perturbation operator in the unperturbed basis are given by
\begin{align}
    \la m | x^4 | n \ra 
    &= \frac{1}{4} \sqrt{n(n-1)(n-2)(n-3)} \, \delta_{m, n-4} \notag \\ &\phantom{=}+ \frac{1}{4} \sqrt{n(n-1)}(4n-2) \delta_{m, n-2} \notag \\
    &\phantom{=}+ \frac{1}{4} (6n^2 + 6n +3) \delta_{m, n} \\
    &\phantom{=}+ \frac{1}{4} \sqrt{(n+1)(n+2)}(4n+6) \delta_{m, n+2} \notag \\
    &\phantom{=}+ \frac{1}{4} \sqrt{(n+1)(n+2)(n+3)(n+4)} \, \delta_{m, n+4} \notag \, ,
\end{align}
displaying a band structure in the HO basis.

\section{Sensitivity of the norm matrix}

\begin{figure}[t]
    \centering
    \includegraphics[width=0.8\columnwidth]{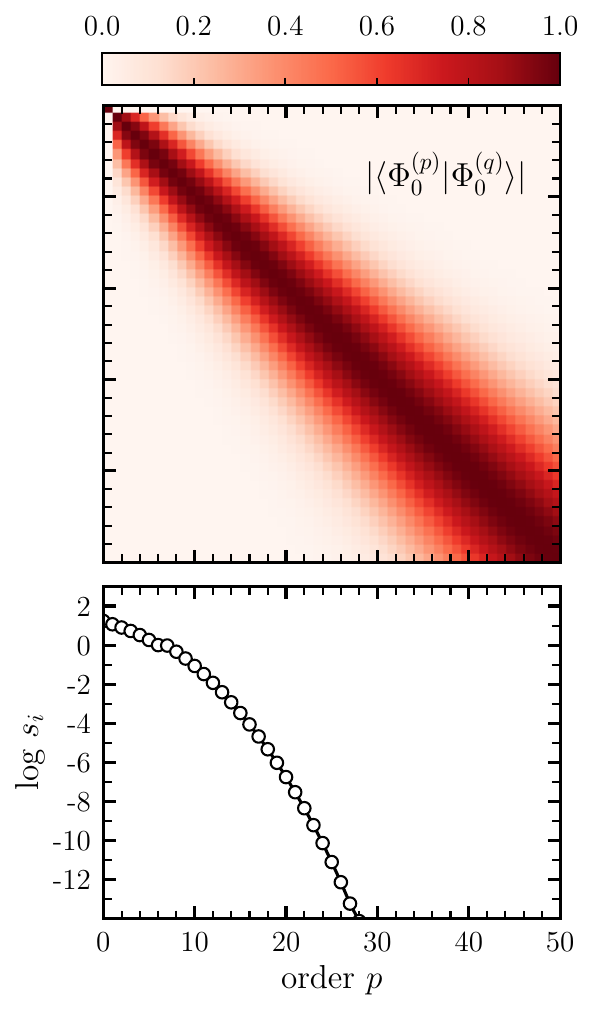}
    \caption{Top panel: Absolute value of norm matrix elements when targeting the unperturbed ground state in PT. 
    Bottom panel: Singular values of the norm matrix arranged in descending order.
    The model space dimension is $\Xmax{n}=50$.}
    \label{fig:norm}
\end{figure}

The quality of the numerical solution of a generalized eigenvalue problem is strongly sensitive to the condition $\kappa$ of the norm matrix,
\begin{align}
    \kappa \equiv \frac{\max s_i}{\min s_i} \, ,
\end{align}
where the $\{s_i\}$ correspond to the non-negative singular values of the norm matrix.
In our cases the norm matrix arises from the overlap of perturbative state corrections that potentially admit (near) linear dependences.

The larger the condition number, the more singular is the norm matrix, such that the presence of (near) linear dependencies make the inversion of the norm matrix numerically challenging.
As a consequence, special care needs to be taken to resolve this at the level of a desired accuracy.
Figure~\ref{fig:norm} shows the norm matrix in the basis of PT state corrections for model-space dimension $\Xmax{n}=50$.
While at low order PT basis vectors point in independent directions, at higher order in the PT expansion linear dependences occur, i.e., $\la \Phi^{(p)}_0 | \Phi^{(p+1)}_0 \ra \approx \pm 1$, such that the norm matrix has large entries near the diagonal.
The presence of such states induces very small singular values of the norm matrix as can be seen from the lower panel of Fig~\ref{fig:norm}.
We observe a rapid falloff of singular values such that $s_i < 10^{-10}$ for $p>25$.
Therefore, high-order state corrections contain a lot of redundant information manifesting in off-diagonal norm matrix entries close to unity (in absolute value) or, equivalently, very small singular values.
In the case of $\Xmax{n}=50$ the observed condition number is given by $\kappa \gtrsim 10^{16}$ hinting at a numerically ill-conditioned problem.

In our calculations this problem was resolved by performing an explicit re-orthogonalization of the PT state-correction basis using Householder transformations yielding a new basis of PT corrections.
Within this basis the norm matrix reduces to the unit matrix and the generalized eigenvalue problem in Eq.~\eqref{eq:geneval} reduces to a standard eigenvalue problem for the transformed Hamiltonian.
Trivially, the condition number is given by $\kappa=1$ in that case.
In principle alternative orthogonalization techniques such as (modified) Gram-Schmidt or QR-decompositions can be applied.
In our applications we did not find a strong sensitivity on the employed method for the range of desired accuracy.

\section{Low-lying spectrum from eigenvector continuation}

\begin{figure}[t]
    \centering
    \includegraphics[width=1.\columnwidth]{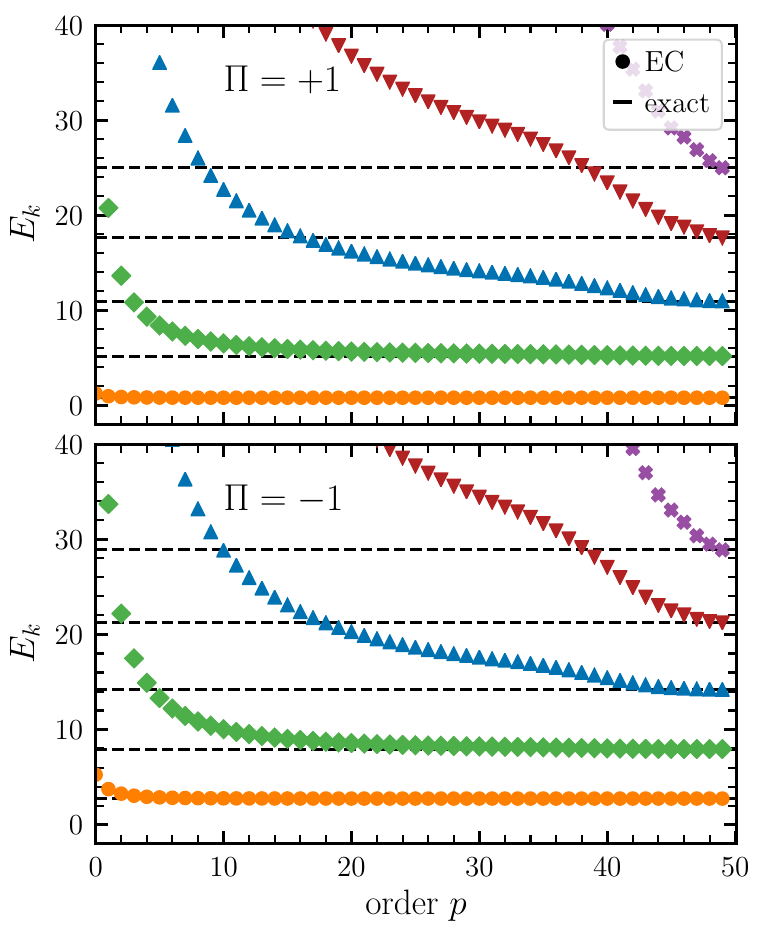}
    \caption{Energies of the low-lying states as a function of EC order. The top (bottom) panel corresponds to states with positive (negative) parity.  The model-space dimension is given by $\Xmax{n}=50$.}
    \label{fig:ECgs}
\end{figure}

Figure~\ref{fig:ECgs} displays the lowest five eigenstates obtained from \ECstar{0} for model-space dimension $\Xmax{n}=50$ in each parity subspace.
The low-lying exact eigenenergies are indicated by black lines.
We observe in both parity subspaces similar convergence properties.
In all cases the exact energy is recovered in a numerically stable way for ground and excited states.
However, the number of EC basis states required to reach convergence changes drastically for different states in the spectrum.
While good accuracy for the ground state is obtained within $p \approx 5$, excited states require larger dimensions, e.g., $p\approx20$ for $k=1$. Note that the subspace probed by $\Mecpt$ at order $p=\Xmax{n}$ is identical to the full basis space such that the EC energies have to coincide with the full diagonalization results. For $k\geq 2$ we observe a significantly slower convergence and some untypical convergence patterns. In particular, already for $k \ge 3$ the full model space needs to be exhausted to reproduce the exact energies.
Nevertheless, convergence is in all cases monotonic as imposed by the underlying variational principle.

\begin{figure}[t!]
    \centering
    \includegraphics[width=1.\columnwidth]{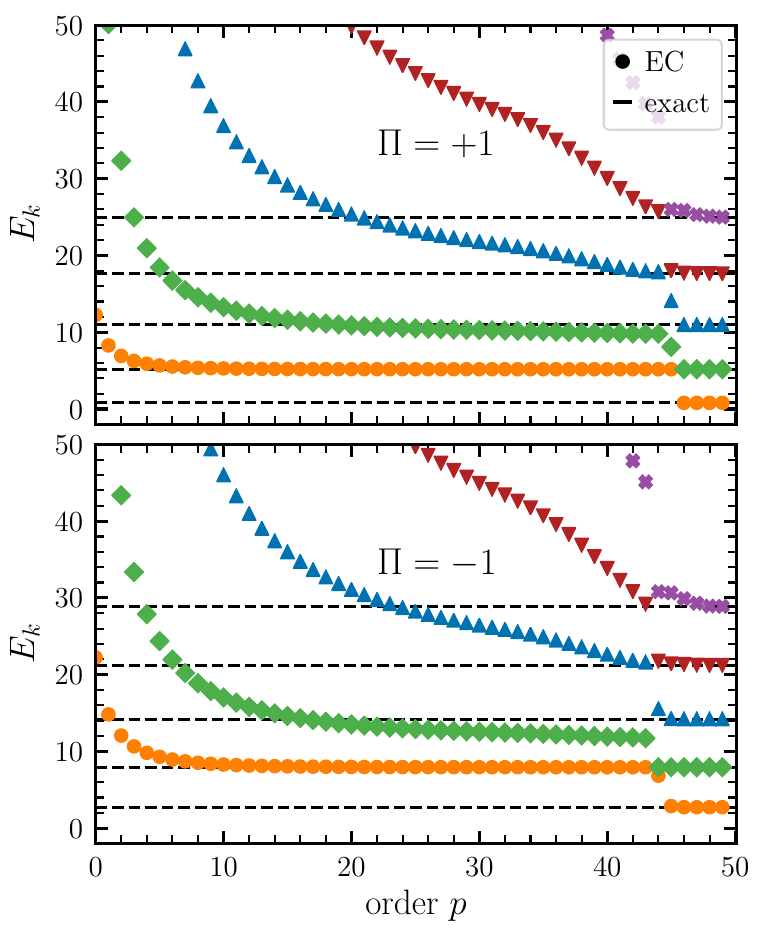}
    \caption{Binding energies from \ECstar{1} using a PT basis constructed from the first excited state in the unperturbed spectrum. The model space dimension is given by $\Xmax{n}=50$.}
    \label{fig:ecex}
\end{figure}

\begin{figure}[t!]
    \centering
    \includegraphics[width=1.\columnwidth]{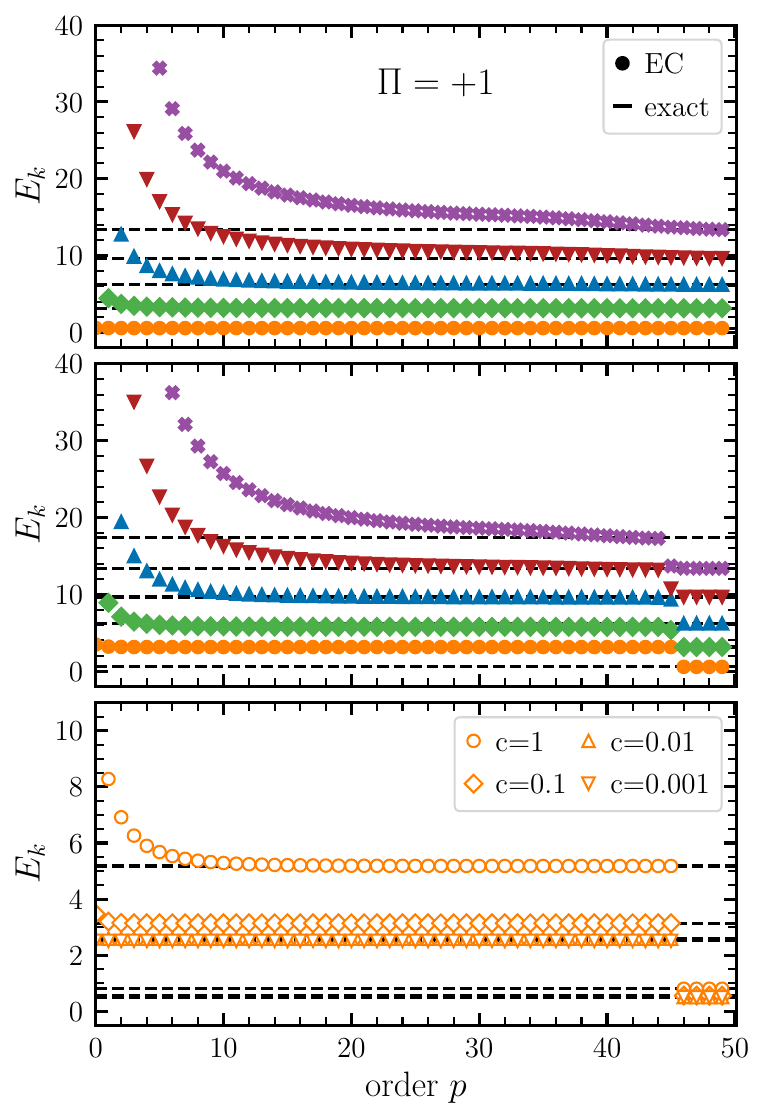}
    \caption{Eigenvector continuation for even parity for targeting the ground state (\ECstar{0}, upper panel) and the first excited state (\ECstar{1}, second panel) for $c=0.1$ and for $c\in\{0.001,0.01,0.1,1\}$ (\ECstar{1}, third panel).}
    \label{fig:01}
\end{figure}

Next the \ECstar{k} framework is benchmarked for $k\neq0$.
Figure~\ref{fig:ecex} shows the low-lying spectrum in both parity blocks obtained from \ECstar{1}, i.e., using the first excited state in the unperturbed spectrum as reference for the subsequent PT treatment.
We observe that for $p\lesssim 40$ the lowest EC state corresponds to the first excited state of the exact spectrum, i.e., the selected subspace does not contain information on the ground-state wave function.
Moreover, in the \ECstar{k}-basis convergence for the $k$-th excited state is most rapid.
This becomes apparent when comparing Figs.~\ref{fig:ECgs} and \ref{fig:ecex}, where the \ECstar{1} convergence of the first excited states is faster than in the case of \ECstar{0}.
This is consistent with the fact that the construction of the basis was optimized for a particular configuration.
We have checked that these observations hold for higher excited states as well.
Eventually, for $p \rightarrow \Xmax{n}$ the EC manifold exhausts the full configuration space.
Consequently, for $p\approx 45$ the EC states become lower in energy since the variational principle enforces the reproduction of the exact spectrum.
At $p=\Xmax{n}$ the full diagonalization energies and results from \ECstar{k} coincide for arbitrary $k$ since EC provides a reparametrization of the same Hilbert space in a transformed basis.
We expect the particular value of $p < \Xmax{n}$ where the EC spectrum starts approaching the ground-state energy to be connected to the particular form of the anharmonic perturbation employed in this work.
While Fig.~\ref{fig:01} shows that these features persist for other couplings (here $c=0.1$), we have not been able to quantitatively predict the position of the jump observed in Fig.~\ref{fig:ecex} and in the lower panel of Fig.~\ref{fig:01}.
Still, we emphasize that the position of the jump is independent of the specific value of $c$.
Since the change of $c$ only results in a rescaling of the perturbative state corrections, the probed subspace for different values of $c$ coincides and, thus, the EC approach will operate on the same manifold of states of the configuration basis.

\section{Comparison of configuration bases}

As discussed previously, the EC framework provides a way of defining a low-dimensional manifold of configurations which is used for a subsequent diagonalization.
Naturally, the quality of the EC results will depend on the quality of the selected subspace for a given observable.
Therefore, we compare the convergence of EC with the canonical CI basis
\begin{align}
    \Mci = \{ | \Phi_k \ra \, : \,  k=1,...,N \} \, 
\end{align}
where $|\Phi_k\ra = \vec{e}_k$ denotes the $k$-th unit vector.
In many-body applications the states $|\Phi_k\ra$ will correspond to simple Slater determinants whereas the PT corrections are of multi-configurational character, i.e., superposition of several Slater determinants.

\begin{figure}[t!]
    \centering
    \includegraphics[width=1.\columnwidth]{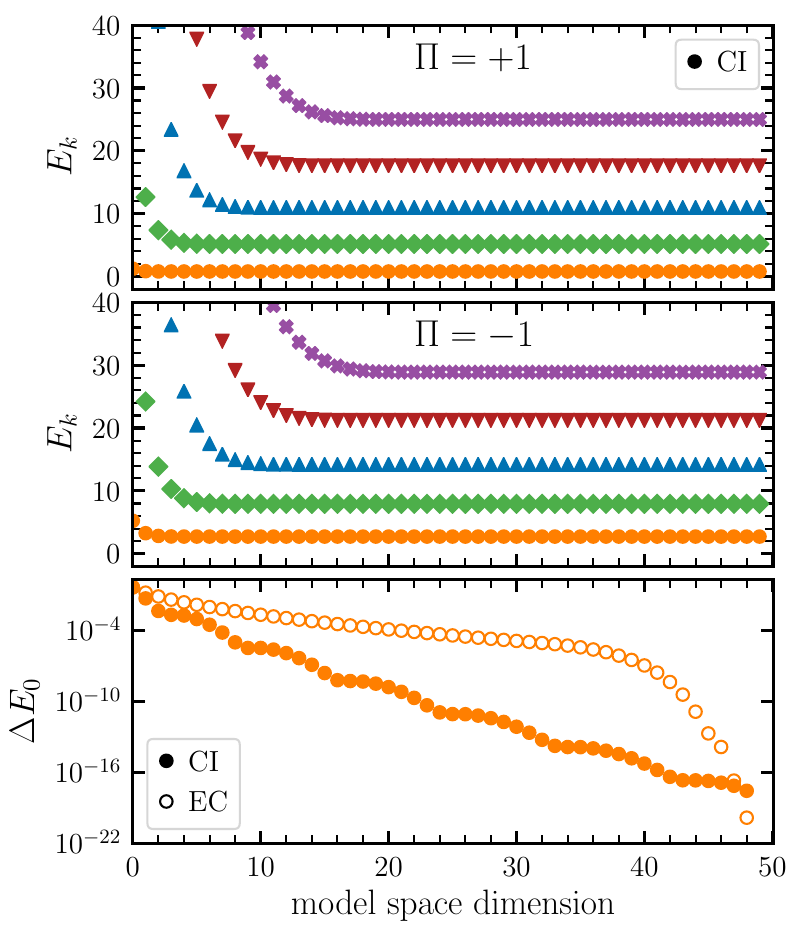}
    \caption{Energies obtained in a standard CI basis as a function of the basis space (top/middle panel). Relative error on ground-state energy from CI and \ECstar{0} calculations for $\Xmax{n}=50$ (bottom panel). For all shown results $c=1$ has been used.}
    \label{fig:cicomparison}
\end{figure}

Figure~\ref{fig:cicomparison} displays the convergence with the model space dimension of the low-lying spectrum in both parity subspaces.
The convergence as a function of basis size is much faster and more natural than in the case of an EC-selected subspace.
In particular, the two lowest states for each parity can be accurately extracted using $p \lesssim 10$. 
Even for the fourth excited state less than 20 configurations are needed to reproduce the full diagonalization result.
Comparing this to the quality of low-order EC calculations indicates that in the present case the canonical basis performs much better than any of the choices in the EC approach.
This is not surprising, since the quality of the EC manifold is directly connected to the quality of the PT state corrections.
Since we are working in a non-perturbative environment, i.e., exponentially divergent perturbation series, PT-based wave functions will not accurately approximate the true eigenstates.
In fact, PT state corrections seem not to pick optimal subspaces of the Hilbert space in our case and, therefore, give slower convergence in EC applications.
This is to be contrasted with previous applications using softer Hamiltonians in many-body problems, where the EC was able to efficiently select the relevant low-dimensional subspace out of a large set of Slater determinants~\cite{Demol2020EC}.

\section{Conclusion}

In this Letter, EC was used to extract the energies of ground and excited states for the case of the anharmonic oscillator with quartic perturbation.
Even in presence of strongly divergent PT expansions the EC-resummed energies robustly converges towards the full diagonalization results where the convergence is fastest for the ground-state expansion.
The convergence of the EC approach for excited states can be improved by using PT state corrections for the excited states themselves, thus, informing the EC basis on the properties of the target wave function.
The \ECstar{0} approach is limited to excited states in the same symmetry class as the fully interacting ground state.
While in aHO applications parity conservation separates the spectrum into two parts, most large-scale \emph{ab initio} applications are performed in a symmetry-restricted way, such that only states with the same total angular-momentum are accessible from such calculations, e.g., low-lying $J^\Pi = 0^+$ states in the case of even-even nuclei.
We note that the design of \ECstar{k} in realistic applications is technically more complicated due to the presence of degeneracies in the unperturbed spectrum making the formulation of PT for excited states more complex.

The EC approach can be seen as a selected CI approach employing a contracted basis from PT-based wave function corrections. 
As such it suffers from the same limitations as truncated CI such as the lack of size-extensivity at finite truncation order.
In the future, we will investigate size-extensivity properties, which will require the evaluation of low-order EC approximations for large system sizes.
These developments will be supported by deriving a diagrammatic expansion of the EC formalism at low order which circumvents the formation of the exponentially large configuration space.

\section*{Acknowledgements}

We thank Dean Lee for helpful discussions.
This work was supported in part by the  Deutsche  Forschungsgemeinschaft  (DFG,  German Research Foundation) -- Projektnummer 279384907 -- SFB 1245, by the European Research Council (ERC) under the European Union's Horizon 2020 research and innovation programme (Grant Agreement No.~101020842), and by the Max Planck Society.

\bibliographystyle{apsrev4-1}
\bibliography{strongint}

\end{document}